\documentstyle[epsf,epsfig]{article}
\makeatletter

          \@addtoreset{equation}{section}
       \makeatother

\newcommand{\be}{\begin{equation}}
\newcommand{\ee}{\end{equation}}
\newcommand{\ba}{\begin{eqnarray}}
\newcommand{\ea}{\end{eqnarray}}
\newcommand{\bd}{\begin{displaymath}}
\newcommand{\ed}{\end{displaymath}}

\begin{document}
\title{%
%\vskip 1cm
\bf 
Renormalization group analysis of the Higgs boson 
mass in a noncommutative differntial geometry
%\vskip 5cm
}
\author{%
Yoshitaka {\sc Okumura}\thanks{
e-mail address: okum@isc.chubu.ac.jp}\\
{\it Department of Natural Science, 
Chubu University, Kasugai, 487, Japan
}}
\date{}
\maketitle
\vskip 0.5cm
\begin{abstract}
{
\normalsize
Within one loop approximation, 
the renormalization group analysis of the Higgs boson mass is performed 
with an initial condition  induced by $m_{_H}=\sqrt{2}m_{_W}$ 
which is presented in the new scheme 
of our noncommutative differential geometry
 for the reconstruction of the standard model. 
This initial condition holds at energy derived from
 $g^2=(5/3)g'^2$ with the running SU(2)$_{_L}$ and
U(1)$_{_Y}$ gauge coupling constants.  $\sin^2\theta_{_W}=3/8$
is obtained under same conditions. 
However, contrary to ${\rm SU(5)}$ GUT without 
supersymmetry, the grand unification of coupling constant is not 
realized in this scheme.
The physical mass of the Higgs boson considerably 
depends on the top quark mass $m_{top}$ since the top quark Yukawa coupling
constants  gives a large negative contribution to  $\beta$ 
function of the Higgs quartic coupling constant in  wide range.
The Higgs boson mass varies from 153.42GeV to 191.94GeV
corresponding to $168{\rm GeV}\leq m_{top}\leq 192{\rm GeV}$. 
We find $m_{_H}=164.01$GeV for $ m_{top}=175$GeV and 
$m_{_H}=171.92$GeV for $ m_{top}=180$GeV.
\vskip 0.2cm
\noindent
PACS number(s): 11.15.Ex,02.40.-k,12.15.-y,12.10.Kt}
\end{abstract}
%\newpage
%%%%%%%%%%%%%%%%%%%%%%%%%%%%%%%%%%%%%%%%%%%%%%%%%%%%%%%%%%%%%%%%%%%%%%%%%%%%%
\thispagestyle{empty}
%%%%%%%%%%%%%%%%%%%%%%%%%%%%%%%%%%%%%%%%%%%%%%%%%%%%%%%%%%%%%%%%%%%%%%%%%%%%%
%\vfill\eject
%%%%%%%%%%%%%%%%%%%%%%%%%%%%%%%%%%%%%%%%%%%%%%%%%%%%%%%%%%%%%%%%%%%%%%%%%%%
\section{Introduction}
Noncommutative differential geometry (NCG) on the discrete space such as
$M_4\times Z_2$ and, in general, $M_4\times Z_{N}(N=2,3,\cdots)$ 
; Minkowski space multiplied by several points space 
allows one to regard the Higgs field
as the gauge field on the same footing as the ordinary gauge field.
A number of articles in various versions of NCG 
\cite{Con}-\cite{Soga} have confirmed this fact
by reconstructing the standard model and the grand unified theory (GUT)
\cite{Cham}, \cite{Osu5}, \cite{O10}
without any theoretical and experimental contradictions. This is 
because the NCG approach to understand the Higgs mechanism 
apparently shows the reason for the existence of the Higgs boson field
and, in addition, it does not 
requires any other extra physical mode except for those in the 
standard model and GUT. \par
The bosonic Lagrangian of the spontaneously broken gauge theory 
derived from the NCG approach has the restrictions on the gauge
and the Higgs quartic coupling constants. These restrictions yield 
the numerical estimate of the Weinberg angle and 
the mass relation between the Higgs boson and other particle
such as gauge boson and top quark, 
{\it though these relations hold only in tree level}.
Several works have been done \cite{AGM}, \cite{SS} 
to estimate the quantum effects of these 
relations by assuming them to hold at some renormalization point. 
Shinohara, Nishida, Tanaka and Sogami \cite{SS} have recently calculated 
in a very clear formulation the renormalization group (RG) evolution 
of the mass relation $m_{H}=\sqrt{2}m_{top}$ for the Higgs boson 
and top quark, which was proposed by Sogami \cite{Soga}
in a variant of NCG with the generalized covariant derivative with 
gauge and Higgs boson fields operating on the total fermion field. \par
Since Connes proposed the first idea \cite{Con} 
concerning NCG, many versions of the
NCG approach \cite{MM}-\cite{Soga} 
have appeared to understand the Higgs mechanism, among which 
the present author \cite{MO1}, \cite{Osu5} 
has also proposed a characteristic formulation of NCG.
Our formulation of a NCG is a  generalization of 
the usual differential geometry on an ordinary manifold to the discrete
manifold $M_4\times Z_{N}$.
The reconstruction of SO(10) GUT \cite{O10} and left-right symmetric
gauge model \cite{Olr} had been already performed according to 
our NCG scheme.
In a NCG on $M_4\times Z_2$,
the extra differential one-form $\chi$ in $Z_2$
is introduced in addition to
the usual one-form $dx^\mu$ in $M_4$, 
and therefore our formulation is very similar
to ordinary differential geometry, 
 whereas, in Connes'original one and 
his follower's NCG, the Dirac matrices $\gamma_\mu$ and $\gamma_5$ 
are used  to describe the generalized gauge field.
In a NCG, the  gauge  and the Higgs boson fields are given as coefficients
of $dx^\mu$ and $\chi$ in the generalized connection on $M_4\times Z_2$, 
respectively. However, there is no symmetry to mix $dx^\mu$ and $\chi$, 
and, therefore, the ordinary gauge field can not be transformed to 
the Higgs boson field.
In Ref.\cite{OFS}, three generations of fermions including 
left and right-handed quarks and leptons and the strong interaction
are incorporated in a new scheme of our NCG to reconstruct the standard 
model. By setting coupling constants on the $y=+$ and
$y=-$ planes with  an argument $y(=\pm)$ in $Z_2$ to be equal, we could
 obtain
$\sin^2\theta_{W}=3/8$ and $m_{H}=\sqrt{2}m_{W}$. 
Following Shinohara and co-workers \cite{SS}, 
we will perform in this article the
RG analysis of the mass relation 
$m_{H}=\sqrt{2}m_{W}$ for the gauge and Higgs bosons, 
which  holds at the energy derived from  $g^2=(5/3)g'^2$ as for the running
SU(2)$_{_L}$ and U(1)$_{_Y}$ gauge coupling constants.
In other version of NCG including Sogami's approach \cite{Soga}, 
the mass relation
between the Higgs boson and top quark is presented, whereas ours 
 is the mass relation between the Higgs and charged gauge bosons.
This is because the Yukawa coupling 
constants written in matrix form in the generation space are
not contained in the generalized gauge field in our formulation.
\par
This article consists of  five sections. The next section presents 
the brief review  of our new scheme of a NCG in Ref.\cite{OFS} which 
incorporates the strong interaction.
In the third section, the reconstruction of the standard model
 based on the new scheme of our NCG presented in the second section
 is reviewed. A characteristic point is to take the fermion field
 to be a vector in 24-dimensional space containing color and three
 generations indices.
The fourth section, the main section in this paper, 
includes the numerical estimation of 
the RG evolution of 
the mass relation $m_{H}=\sqrt{2}m_{W}$ on the assumption that
it holds at a energy of 
$g^2=(5/3)g'^2$ with SU(2)$_{_L}$ and U(1) gauge coupling constants. 
The last section is devoted to concluding remarks.
%%%%%%%%%%%%%%%%%%%%%%%%%%%%%%%%%%%%%%%%%%%%%%%%%%%%%%%%%%%%%%%%%%%%%%%%%%%%%%
\section{ Our NCG on $M_4\times Z_2$}
We review our previous formulation \cite{OFS}
proposed to reconstruct the standard model  in which 
three generations of left and right-handed quarks and leptons
are taken into account and also the  strong interaction is incorporated.
A characteristic point of this formulation is to take the left 
and right-handed fermions $\psi(x,y)$ with arguments $x$ 
and $y(=\pm) $ in $M_4$ and $Z_2$, respectively as
\begin{equation}
       \psi(x,+)=\left(\matrix{ 
                                u^r_{_{^L}}\cr
                                u^g_{_{^L}}\cr
                                u^b_{_{^L}}\cr
                                \nu_{_{^L}}\cr
                                d^r_{_{^L}}\cr
                                d^g_{_{^L}}\cr
                                d^b_{_{^L}}\cr
                                e_{_{^L}}\cr }
                            \right), 
\hskip 1.0cm
       \psi(x,-)=\left(\matrix{ 
                                u^r_{_{^R}}\cr
                                u^g_{_{^R}}\cr
                                u^b_{_{^R}}\cr
                                 0       \cr
                                d^r_{_{^R}}\cr
                                d^g_{_{^R}}\cr
                                d^b_{_{^R}}\cr
                                e_{_{^R}}\cr }
                            \right), \label{2.1}
\end{equation}
where subscripts $L$ and $R$ denote the left-handed and
right-handed fermions, respectively and 
superscripts $r$, $g$ and $b$ represent the color indices. 
It should be noticed that $\psi(x,y)$ has the index for the three 
generation and so do the explicit expressions 
for fermions on the right hand sides
of Eq.(\ref{2.1}).  In the strict expressions,
 $u,$ $d,$  $\nu$ and $e$
 in  Eq.(\ref{2.1}) should be replaced by
\begin{equation}
       u \to \left(\matrix{ u \cr
                            c \cr
                            t\cr}
                            \right), 
                           \hskip 0.2cm
       d \to \left(\matrix{ d \cr
                            s \cr
                            b \cr}
                            \right), \hskip 0.2cm
       \nu \to \left(\matrix{ \nu_e  \cr
                              \nu_\mu \cr
                              \nu_\tau\cr}
                            \right), \hskip 0.2cm
       e \to \left(\matrix{ e  \cr
                            \mu \cr
                            \tau\cr}
                            \right), 
                            \label{2.2}
\end{equation}
respectively. Thus, $\psi(x,\pm)$ is a vector in the 24-dimensional space.
In order to construct the Dirac Lagrangian of standard model
corresponding to $\psi(x,\pm)$ in Eqs.(\ref{2.1}), 
we must need 24-dimensional generalized 
 covariant derivative composed of gauge and
Higgs fields on $M_4\times Z_2$. This is achieved by developing
a NCG on the discrete space in the follows.
\par
Let us start with the equation of the generalized gauge field 
${\cal A}(x,y)$
written in one-form on the discrete space $M_4\times Z_2$.
\begin{equation}
      {\cal A}(x,y)=\sum_{i}a^\dagger_{i}(x,y){\bf d}a_i(x,y)+
      \sum_{j}b^\dagger_{j}(x,y){\bf d}b_j(x,y),\label{2.3}
\end{equation}
where 
$a_i(x,y)$ and $b_j(x,y)$ are the square-matrix-valued functions
and  are taken so as to commute with each other, because
$\sum_{i}a^\dagger_{i}(x,y){\bf d}a_i(x,y)$ is 
the flavor sector including the flavor gauge and the Higgs fields 
and $\sum_{j}b^\dagger_{j}(x,y){\bf d}b_j(x,y)$ correspond with 
 color gauge field. 
$i$ and $j$ are  variables of the extra
internal space which we can not now identify. 
${\bf d}$ in Eq. (\ref{2.3}) is the generalized exterior 
derivative defined as follows.
\begin{eqnarray} 
&&       {\bf d}=d + d_\chi , \label{2.4}\\  
&&     da_i(x,y) = \partial_\mu a_i(x,y)dx^\mu,\hskip 1cm \label{2.5}\\
&&   d_{\chi} a_i(x,y) =[-a_i(x,y)M(y) + M(y)a_i(x,-y)]\chi,
        \label{2.6}\\
&&      {d}b_j(x,y)= \partial_\mu b_j(x,y)dx^\mu \label{2.8}\\
&&   d_{\chi} b_j(x,y) =0. \label{2.9}
\end{eqnarray}
Here 
$dx^\mu$ is 
ordinary one-form basis, taken to be dimensionless, in Minkowski space 
$M_4$, and $\chi$ 
is the one-form basis, assumed to be also dimensionless, 
in the discrete space $Z_2$.
If the operator $\partial_y$ is defined as
\begin{equation}
    \partial_ya_i(x,y)=[-a_i(x,y)M(y) + M(y)a_i(x,-y)],  \label{2.6a}
\end{equation}
Eq.(\ref{2.6}) is rewritten as
\begin{equation}
    d_{\chi} a_i(x,y) =\partial_ya_i(x,y)\chi,   \label{2.6b}
\end{equation}
just in the same form as in Eq.(\ref{2.5}).
The operator $\partial_y$ is a difference operator on the discrete space
with the modification by $M(y)$. $M(y)$ in Eq.(\ref{2.6a}) 
is required because $a_i(x,+)$ and $a_i(x,-)$ are in general 
square matrices with different degree and, therefore, $M(y)$ ensures the
consistent calculations of matrices. 
$\partial_y$ in $Z_2$ is an alternative of $\partial_\mu$ in $M_4$.
We have introduced $x$-independent matrix $M(y)$ 
whose hermitian conjugation is given by $M(y)^\dagger=M(-y)$. 
The matrix $M(y)$ determines the scale and pattern of 
the spontaneous breakdown of the gauge symmetry. Thus, 
Eq.(\ref{2.9}) means that the color symmetry 
of the strong interaction does not break spontaneously.\par
In order to find the explicit forms of gauge and Higgs fields
according to Eq.(\ref{2.3}), we need the following 
important algebraic rule of non-commutative geometry.
\begin{equation}
        f(x,y)\chi=\chi f(x,-y), \label{2.10}
\end{equation}
where $f(x,y)$ is a field defined on the discrete space such as
$a_i(x,y)$, gauge field, Higgs field or fermion fields.
It should be noticed that Eq.(\ref{2.10}) does not express 
the relation between
the matrix elements of $f(x,+)$ and $f(x,-)$ but insures the product between
the fields expressed in  differential form on the discrete space. 
Using Eq.(\ref{2.10}) and some other algebraic rules 
in Eqs.(\ref{2.4})-(\ref{2.9}),
${\cal A}(x,y)$ is rewritten as
\begin{equation}
 {\cal A}(x,y)=A_\mu(x,y)dx^\mu+{\mit\Phi}(x,y)\chi+G_\mu(x)dx^\mu, 
 \label{2.11}
\end{equation}
where
\begin{eqnarray}
&&    A_\mu(x,y) = \sum_{i}a_{i}^\dagger(x,y)\partial_\mu a_{i}(x,y), 
                                   \label{2.12}\\
&&     {\mit\Phi}(x,y) = \sum_{i}a_{i}^\dagger(x,y)\,(-a_i(x,y)M(y) 
            + M(y)a_i(x,-y))  \nonumber\\
&&\hskip 1.1cm  =a_{i}^\dagger(x,y)\,\partial_ya_i(x,y), \label{2.13}\\
&&  G_\mu(x)=\sum_{j}b_{j}^\dagger(x)\partial_\mu b_{j}(x).
  \label{2.14}
\end{eqnarray}
$A_\mu(x,y)$, ${\mit\Phi}(x,y)$ and $G_\mu(x)$ are identified with
the gauge field in the flavor symmetry, Higgs fields,
and the color gauge field responsible for the strong interaction,
respectively. 
\par
In order to identify  $A_\mu(x,y)$ and $G^{}_\mu(x)$ 
with true gauge fields, 
the following conditions have to be imposed.
\begin{eqnarray}
&&    \sum_{i}a_{i}^\dagger(x,y)a_{i}(x,y)= 1,  \label{2.15}\\
&&     \sum_{j}b_{j}^\dagger(x)b_{j}(x)={1\over g_s},      
  \label{2.16}
\end{eqnarray}
where $g_s$ is a constant related to 
the  coupling constant of the strong interaction. 
\par
Before constructing the gauge covariant field strength, 
we address the gauge transformation 
of $a_i(x,y)$ and $b_j(x)$ which is defined as 
\begin{eqnarray}
&&      a^{g}_{i}(x,y)= a_{i}(x,y)g_{f}(x,y), \nonumber\\
&&      b^{g}_{j}(x) =  b_j(x)g_c(x),     
\label{2.17}
\end{eqnarray}
where
$g_{f}(x,y)$ and $g_c(x)$ are the gauge functions 
with respect to the flavor unitary group 
and the color SU(3)$_c$ group, respectively. 
It should be noticed that $g_c(x)$ can be taken 
to commute with $a_i(x,y)$ and 
$M(y)$ and at the same time $g_{f}(x,y)$ is taken 
to commute with $b_j(x)$. 
$g_{f}(x,y)$ and $g_c(x)$ commute with each other.
Then, we obtain the gauge transformation of ${\cal A}(x,y)$ to be
\begin{equation}
{\cal A}^g(x,y)=g^{-1}_{f}(x,y) 
g_c^{-1}(x){\cal A}(x,y)g_{f}(x,y)g_c(x)
+g^{-1}_{f}(x,y){\bf d}g_{f}(x,y)+ {1\over g_s}
\,g^{-1}_c(x){d}g_c(x), \label{2.18}
\end{equation}
where use has been made of Eq.(\ref{2.3}) and Eq. (\ref{2.17}),
and  as in Eqs.(\ref{2.5}) and (\ref{2.6}),
\begin{equation} 
       {\bf d}g_{f}(x,y)=\partial_\mu g_{f}(x,y)dx^\mu
       +\partial_yg_{f}(x,y)\chi
          \label{2.19}
\end{equation}
with the operator $\partial_y$ defined in Eq.(\ref{2.6a}).
Using Eqs.(\ref{2.18}) and (\ref{2.19}), 
we can find the gauge transformations of gauge and Higgs fields as
\begin{eqnarray}
&&       A_\mu^g(x,y)=g^{-1}_{f}(x,y)A_\mu(x,y)g_{f}(x,y)
                          +  g^{-1}_{f}(x,y)\partial_\mu g_{f}(x,y),  
                           \label{2.20}\\
&&       {\mit\Phi}^g(x,y)=g^{-1}_{f}(x,y){\mit\Phi}(x,y)g_{f}(x,-y)
             +    g^{-1}_{f}(x,y)\partial_y g_{f}(x,y), 
                             \label{2.21}\\
&&      G_\mu^g(x)=g^{-1}_c(x)G_\mu(x)g_c(x)+
                           \frac1{g_s}g^{-1}_c(x)\partial_\mu g_c(x).  
                           \label{2.22}
\end{eqnarray}
Equation(\ref{2.21}) is very similar to two other  equations and so 
it strongly indicates that the Higgs field is a kind of gauge field
on the discrete space. From Eqs.(\ref{2.19}) and 
(\ref{2.21}) it is rewritten as
\begin{equation}
       {\mit\Phi}^g(x,y)+M(y)=g^{-1}_{f}(x,y)({\mit\Phi}(x,y)
               +M(y))g_{f}(x,-y), \label{2.23}\\
\end{equation}
which makes it obvious that 
\begin{equation}
H(x,y)={\mit\Phi}(x,y)+M(y) \label{2.24a}
\end{equation}
is un-shifted Higgs field whereas ${\mit\Phi}(x,y)$ denotes shifted one with
the vanishing vacuum expectation value.\par
In addition to the algebraic rules in Eqs.(\ref{2.4})-(\ref{2.9})
 we add one more important rule that
\begin{equation}
              d_\chi M(y)=M(y)M(-y)\chi            \label{2.24}
\end{equation}
which yields together with Eqs.(\ref{2.4})-(\ref{2.9}) the nilpotency 
of the generalized exterior derivative $\bf d$;
\begin{equation}
          {\bf d}^2 f(x,y)=(d^2+dd_\chi+d_\chi d
          +d_\chi^2)f(x,y)=0  \label{2.25}
\end{equation}
with the condition $dx^\mu\wedge\chi=-\chi\wedge dx^\mu$.
In proving the nilpotency,  the following rule must be taken into account 
that whenever the $ d_\chi$ operation jumps over $M(y)$, minus sign 
is attached, for example
\begin{eqnarray}
  && d_\chi \{a(x,y)M(y)b(x,-y)\}=(d_\chi a(x,y))M(y)b(x,-y)\nonumber\\
                 &&  \hskip5cm+ a(x,y)(d_\chi M(y))b(x,-y) 
                         -a(x,y)M(y)(d_\chi b(x,-y)).
                                        \label{2.25a}
\end{eqnarray}
With these considerations we can construct the gauge covariant field
strength.
\begin{equation}
  {\cal F}(x,y)= F(x,y) +  {\cal G}(x),
\label{2.26}
\end{equation}
where $F(x,y)$ and ${\cal G}(x)$ are the field strengths 
of flavor and color gauge fields, respectively and given as
\begin{eqnarray}
&&     F(x,y) = {\bf d}A(x,y)+A(x,y)\wedge A(x,y),     \nonumber\\
&&     {\cal G}(x)   =d\,G(x)+g_s G(x)\wedge G(x), 
\label{2.27}
\end{eqnarray}
where it should be noted that 
${\bf d}A(x,y)=\sum_i{\bf d}a_i^\dagger(x,y)\wedge {\bf d}a_i(x,y)$ 
and $d\,G(x)=\sum_j{ d}b_j^\dagger(x)\wedge {d}b_j(x)$  
are followed because of the nilpotency of $\bf d$ and $d$. 
We can easily find the gauge 
transformation of ${\cal F}(x,y)$ as
\begin{equation}
         {\cal F}^g(x,y)=g^{-1}(x,y){\cal F}(x,y)g(x,y),  \label{2.28}
\end{equation}
where $g(x,y)=g_{f}(x,y)g_c(x)$.
The algebraic rules defined in Eqs.(\ref{2.4})-(\ref{2.10})
and (\ref{2.24}) yield
\begin{eqnarray}
 F(x,y ) &=& { 1 \over 2}F_{\mu\nu}(x,y)dx^\mu \wedge dx^\nu 
             + D_\mu {\mit\Phi}(x,y)dx^\mu \wedge \chi 
               + V(x,y)\chi \wedge \chi,
                \label{2.29}
\end{eqnarray}
where
\begin{eqnarray}
 && F_{\mu\nu}(x,y)=\partial_\mu A_\nu (x,y) - \partial_\nu A_\mu (x,y) 
                +[A_\mu(x,y), A_\mu(x,y)],\label{2.30}\\
 && D_\mu {\mit\Phi}(x,y)=\partial_\mu {\mit\Phi}(x,y)
    + A_\mu(x,y)(M(y) + {\mit\Phi}(x,y))
              -({\mit\Phi}(x,y)+M(y))A_\mu(x,-y),\label{2.31}\\
&&  V(x,y)= ({\mit\Phi}(x,y) + M(y))({\mit\Phi}(x,-y) 
               + M(-y)) - Y(x,y). \label{2.32}
\end{eqnarray}
$Y(x,y)$ in Eq.(\ref{2.32}) is auxiliary field and expressed as 
\begin{equation}
  Y(x,y)= \sum_{i}a_{i}^\dagger(x,y)M(y)M(-y)a_{i}(x,y),
 \label{2.33}
\end{equation}
which may be independent or dependent of ${\mit\Phi}(x,y)$ 
and/or may be a constant field.
In contrast to $F(x,y)$, ${\cal G}(x)$ is simply denoted as
\begin{eqnarray}   
 {\cal G}(x)&=&{1\over 2}{G}_{\mu\nu}(x)dx^\mu\wedge dx^\nu \nonumber\\
        &=&{1\over 2}\{\partial_\mu G^{}_\nu(x)-\partial_\nu G^{}_\mu(x)
     + g_s[G^{}_\mu(x), G^{}_\mu(x)]\}dx^\mu\wedge dx^\nu. 
\label{2.34}
\end{eqnarray}
\par
With the same metric structure on the discrete space 
$M_4\times Z_{2}$ as in Ref.\cite{MO1} that
\begin{eqnarray}
&& <dx^\mu, dx^\nu>=g^{\mu\nu},\quad 
g^{\mu\nu}={\rm diag}(1,-1,-1,-1),\nonumber\\
&& <\chi, dx^\mu>=0,\nonumber\\
&& <\chi, \chi>=-1
\label{2.35}
\end{eqnarray}
 we can obtain the gauge invariant 
Yang-Mills-Higgs lagrangian(YMH)
\begin{eqnarray}
{\cal L}_{_{YMH}}(x)&=&-{\rm Tr}\sum_{y=\pm}{1 \over g_{y}^2}
< {\cal F}(x,y),  {\cal F}(x,y)>\nonumber\\
&=&-{\rm Tr}\sum_{y=\pm}{1\over 2g^2_y}
F_{\mu\nu}^{\dag}(x,y)F^{\mu\nu}(x,y)\nonumber\\
&&+{\rm Tr}\sum_{y=\pm}{1\over g_{y}^2}
    (D_\mu {\mit\Phi}(x,y))^{\dag}D^\mu {\mit\Phi}(x,y)  \nonumber\\
&& -{\rm Tr}\sum_{y=\pm}{1\over g_{y}^2}
        V^{\dag}(x,y)V(x,y)  \nonumber\\
&&-{\rm Tr}\sum_{y=\pm}{1\over 2g_{y}^2}{ G}_{\mu\nu}^{\dag}(x)
{ G}^{\mu\nu}(x),
\label{2.36}
\end{eqnarray}
where $g_y$ is a constant relating 
to the coupling constant of the flavor gauge field and
Tr denotes the trace over internal symmetry matrices including the color, 
flavor symmetries and generation space. 
The third term on the right hand side 
is the potential term of Higgs particle.
\par
The fermion sector of the standard model was also reconstructed 
in Ref.\cite{OFS}. However, it is not presented in this article 
because it is not necessary to perform the RG analysis.
With these preparations, we can apply the formulation 
in this section to reconstruction of the Yang-Mills-Higgs Lagrangian
in the standard model.

%%%%%%%%%%%%%%%%%%%%%%%%%%%%%%%%%%%%%%%%%%%%%%%%%%%%%%%%%%%%%%%%%%%%%%%%%%%
\section{ Reconstruction of standard model}
According to a NCG in the previous section, the standard model is
reconstructed. This section is also the review of Ref.\cite{OFS}.
The gauge fields $A_\mu(x,y)$ and $G_\mu(x)$
in the covariant derivative must be the differential representation of 
the fermion fields in Eqs.(\ref{2.1}) and (\ref{2.2}) and, therefore,
they are expressed in $24\times24$ matrices. 
The Higgs field ${\mit\Phi}(x,y)$ is also taken 
to give the correct Yukawa interaction in the Dirac Lagrangian
and expressed in the $24\times24$ matrix.
We specify   ${A}_\mu(x,y),$ ${\mit\Phi}(x,y)$ and $G_\mu(x)$ 
in Eq.(\ref{2.11}) as follows:
\begin{equation}
      A_\mu(x,+)=-\frac i2\left\{\sum_{k=1}^3\tau^k\otimes1^4 {A_{L}^k}_\mu
                          + aB_\mu\right\}\otimes 1^3,  
            \label{3.1} 
\end{equation}
where
${A_{L}^k}_\mu$ and $B_\mu$ are $SU(2)_{L}$ 
and U(1) gauge fields, respectively 
and so $\tau^k$ is the Pauli matrices.
$1^3$ represents the unit matrix in the generation space and
$a$ is the U(1) hypercharge matrix
corresponding to $\psi(x,+)$ in Eq.(\ref{2.1}) and expressed as
\begin{equation}
     a={\rm diag}\;(\frac13,\frac13,\frac13,-1,
     \frac13,\frac13,\frac13,-1).     \label{3.2}
\end{equation}
\begin{equation}
      A_\mu(x,-)=-\frac i2
                          bB_\mu\otimes 1^3,  
            \label{3.3} 
\end{equation}
where
$b$ is the U(1) hypercharge matrix
corresponding to $\psi(x,-)$ in Eq.(\ref{2.1}) and so it is
$8\times8$ diagonal matrix
expressed in 
\begin{equation}
     b={\rm diag}\;(\frac43,\frac43,\frac43,0,-\frac23,-\frac23,-\frac23,-2).
     \label{3.4}
\end{equation}
$G_\mu(x)$ is denoted by
\begin{equation}
       G_\mu(x)=-\frac i2\sum_{a=1}^8 \tau^0\otimes 
             \lambda'^a G_\mu^a\otimes 1^3,  \label{3.5}
\end{equation}
where $\lambda'^a$ is $4\times 4$ matrix made of the Gell-Mann 
matrix $\lambda^a$ by adding $0$ components 
to fourth line and column.
\begin{equation}
   \lambda'^a  =\left(\matrix{  & &  & 0\cr
                                & \lambda^a &  & 0\cr
                                &  &   &0\cr
                              0 & 0 & 0 & 0 \cr}
                      \right).     \label{3.6}
\end{equation}
Higgs field ${\mit\Phi}(x,y)$ is represented in $24\times24$ matrix by
\begin{eqnarray}
   && {\mit\Phi}(+)=\left(\matrix{ \phi^\ast_0 & \phi^+ \cr
                           -\phi^- &   \phi_0  \cr } \right) 
                           \otimes 1^4\otimes 1^3,
                           \nonumber\\
  &&  {\mit\Phi}(-)=\left(\matrix{ \phi_0 & -\phi^+ \cr
                           \phi^- &   \phi_0^\ast  \cr } \right) 
                           \otimes 1^4 \otimes 1^3. 
                           \label{3.7}
\end{eqnarray}
Corresponding to Eq.(\ref{3.7}), symmetry breaking function
$M(y)$ is given by
\begin{equation}
    M(+)=\left(\matrix{   \mu & 0 \cr
                           0 &  \mu  \cr } \right) 
                           \otimes 1^4\otimes 1^3,
\hskip 0.2cm M(-)=M(+)^\dagger.
                           \label{3.8}
\end{equation}
It should be noted that $G_\mu$ is taken so as to commute with
$A_\mu(y)$ and ${\mit\Phi}(y)$.
With these specifications, all quantities needed 
to express the explicit expression of  ${\cal F}(x,y)$ 
in Eq.(\ref{2.26}) can be explicitly written down as follows.
\begin{eqnarray}
&&  {F}_{\mu\nu}(x,+)
            =-\frac i2\left\{\sum_{k=1}^3\tau^k\otimes 1^4 
       {F_{L}^k}_{\mu\nu}
       +\tau^0\otimes a B_{\mu\nu}\right\}\otimes1^3,
         \label{3.9} \\
&&  {F}_{\mu\nu}(x,-)=-\frac i2 b B_{\mu\nu}\otimes1^3, \label{3.10} \\
&&  {G}_{\mu\nu}(x)=-\frac i2\sum_{a=1}^8 \tau^0\otimes 
              \lambda'^aG^a_{\mu\nu}\otimes 1^3, \label{3.11} 
\end{eqnarray}
where   
\begin{eqnarray}
 &&     {F_{L}^k}_{\mu\nu}=
    \partial_\mu {A_{L}^k}_\nu-\partial_\nu {A_{L}^k}_\mu
    +\epsilon^{klm} {A_{L}^l}_\mu {A_{L}^m}_\nu, \label{3.12} \\
&&    B_{\mu\nu}=\partial_\mu B_\nu-\partial_\nu B_\mu,   
     \label{3.13} \\
&&   G_{\mu\nu}^a=\partial_\mu G_\nu^a-\partial_\nu G_\mu^a 
                  +g_sf^{abc}G_\mu^bG_\nu^c. \label{3.14}
\end{eqnarray}

$D_\mu{\mit\Phi}(x,y)$ in Eq.(\ref{2.31}) is represented in
\begin{eqnarray}
      && D_\mu{\mit\Phi}(+)=(D_\mu{\mit\Phi}(-))^\dagger
       =\left\{\partial_\mu h' -\frac i2(
       \sum_{k=1}^3\tau^k {A_{L}^k}_\mu h'+ h'c B_\mu)\right\}\otimes 1^4
       \otimes1^3,
       \label{3.15}
\end{eqnarray}
where $h'$ and $c$ are given as
\begin{equation}
        h'=\left(\matrix{ \phi^\ast_0+\mu & \phi^+ \cr
                           -\phi^- &   \phi_0+\mu  \cr } \right),
          \hskip 1cm
        c=\left(\matrix{ -1 & 0 \cr
                          0 & 1  \cr } \right). \label{3.16}
\end{equation}
The matrix $c$ comes from 
\begin{equation}
          \tau^0\otimes a-b=  
          \left(\matrix{ -1 & 0 \cr
                          0 & 1  \cr } \right)\otimes1^4 
                          =c\otimes 1^4 \label{3.17}
\end{equation}
to insure that Higgs doublet $h=(\phi^+,\phi_0+\mu)^t$ has 
plus one  hypercharge  and ${\tilde h}=i\tau^2h^\ast$ minus one.
$V(x,y)$ in Eq.(\ref{2.32}) is expressed in
\begin{eqnarray}
   &&     V(x,+)=(h'h'^\dagger-\mu^2)\otimes 1^4\otimes1^3, \nonumber\\
   &&     V(x,-)=(h'^\dagger h'-\mu^2)\otimes 1^4\otimes1^3.\label{3.18}
\end{eqnarray}
It should be noticed that $Y(x,y)$ in Eq.(\ref{2.33}) can be estimated
by use of Eq.(\ref{3.8}) to be
\begin{eqnarray}
    Y(x,\pm)&&=\sum_ia_i^\dagger(x,\pm)M(\pm)M(\mp)a_i(x,\pm)\nonumber\\
    &&=\mu^2\sum_ia_i^\dagger(x,\pm)a_i(x,\pm)=\mu^2 1^{24}, 
    \label{3.19}
\end{eqnarray}
where use has bend{equation}n made of Eq.(\ref{2.15}).
\par
Putting above equations into Eq.(\ref{2.36}) 
and rescaling  gauge and Higgs fields 
we can obtain Yang-Mills-Higgs lagrangian for the standard model 
as follows:
\begin{eqnarray}
{\cal L}_{{YMH}}&=&
   -\frac14\sum_{k=1}^3\left(F_{\mu\nu}^k\right)^2 
   -\frac14B_{\mu\nu}^2 \nonumber\\
  &&  +|D_\mu h|^2   -\lambda(h^\dagger h-{\mu}^2)^2 \nonumber\\
 && - \frac{1}{4}
      \sum_{a=1}^8{G^a_{\mu\nu}}^{\dagger}{G^a}^{\mu\nu}, \label{3.20}
\end{eqnarray}
where
\begin{eqnarray}
   &&   F_{\mu\nu}^k=\partial_\mu A_\nu^k-\partial_\nu A_\mu^k
         +g\epsilon^{klm}A_\mu^lA_\nu^m,  \label{3.21}\\
   &&   B_{\mu\nu}=\partial_\mu B_\nu-\partial_\nu B_\mu,\label{3.22}\\
   &&     D^\mu h=[\,\partial_\mu-{i\over 2}\,(\sum_k\tau^kg{A^k_{L}}_\mu
          + \,\tau^0\,g'B_\mu\,)\,]\,h, \hskip 1cm
                   h=\left(\matrix{ \phi^+ \cr
                                      \phi_0+\mu  \cr } \right),  
                                             \label{3.23}\\
   &&   G_{\mu\nu}^a=\partial_\mu G_\nu^a-\partial_\nu G_\mu^a
         +g_cf^{abc}G_\mu^bG_\nu^c, \label{3.24} 
\end{eqnarray}
with
\begin{eqnarray}
  &&     g^2=\frac{g_+^2}{12}, \label{3.25a}\\
  &&    {g'}^2=\frac{2g_+^2g_-^2}{3g_-^2{\rm Tr} a^2+
      3g_+^2{\rm Tr}b^2 }=\frac{g_+^2g_-^2}{16g_+^2+4g_-^2}, \label{3.25}\\
 &&     \lambda=\frac{g_+^2g_-^2}
         {24(g_+^2+g_-^2)}, 
         \label{3.26}\\
 &&        g_c^2=g_s^2\frac{g_+^2g_-^2}{6(g_+^2+g_-^2)}. \label{3.27}
\end{eqnarray}
Equation(\ref{3.25}) yields the Weinberg angle with the parameter 
$\delta={g_+}/{g_-}$ to be
\begin{equation}
          \sin^2\theta_{W}=\frac{3}{4(\delta^2+1)}. \label{3.28}
\end{equation}
The gauge transformation affords  the Higgs doublet $h$ to take the form
that
\begin{equation}
             h=\frac1{\sqrt 2}\left(\matrix{ 0 \cr
                              \eta+v  \cr } \right)  \label{3.29}
\end{equation}
which makes possible along with Eqs.(\ref{3.20})$\sim$(\ref{3.23})
to expect the gauge boson and Higgs particle masses. 
\begin{eqnarray}
 &&       m_{W}^2=\frac1{48}g_+^2v^2, \label{3.30}\\
 &&       m_{Z}^2=\frac{1+\delta^2}{12(4\delta^2+1)}g_+^2v^2, 
 \label{3.31}\\
 &&       m_{H}^2= \frac1{12(\delta^2+1)}g_+^2v^2.    
        \label{3.32}
\end{eqnarray}
In the limit of equal coupling constant $g_+=g_-$, 
 $g^2=(5/3)g'^2$ is followed and, therefore, $\sin^2\theta_{W}=3/8$
which is the same value expected by the $SU(5)$ and $SO(10)$
GUTs. This fact makes possible to conclude that the limit of $g_+=g_-$
 yields the relations that hold at 
 the energy derived from  $g^2=(5/3)g'^2$ with the running SU(2)$_{_L}$ and
 U(1)$_{_Y}$ gauge coupling constants.
 However, it should be noted that 
 the color gauge coupling constant $g_c$ includes
 an additional parameter $g_s$ as in Eq.(\ref{3.27}), and so 
 $g_c$ does not necessarily coincide with the SU(2)$_{_L}$ 
 gauge coupling constant $g$. 
 Thus, contrary to SU(5) GUT without supersymmetry , 
 the grand unification of coupling constants is not realized in the present 
 scheme.   
In the next section, in order to predict the physical 
Higgs mass, we will perform the RG analysis 
 with the initial condition induced by $m_{H}=\sqrt{2}m_{W}$ 
at the energy derived from  $g^2(t)=(5/3)g'^2(t)$ with the running coupling
constans.
%%%%%%%%%%%%%%%%%%%%%%%%%%%%%%%%%%%%%%%%%%%%%%%%%%%%%%%%%%%%%%
%%%%%%%%%%%%%%%%%%%%%%%%%%%%%%%%%%%%%%%%%%%%%%%%%%%%%%%%%%%%%%%%%
\section{Numerical estimation of the Higgs mass}
In order to perform the RG analysis, 
the renormalizable Lagrangian is necessary. We have the Yang-Mills-Higgs
Lagrangian of the standard model expressed in Eq.(\ref{3.20}).
However, it has the special restriction on the gauge couplings 
and the Higgs quartic coupling which yield the 
mass relation for the Higgs and gauge bosons. 
In a sense, the restriction on coupling constants is an obstacle 
for renormalizability.
Thus, we consider that the NCG approach provides the Lagrangian
at some renormalization point. The more general Lagrangian is
obtained by adding the term for the Higgs potential
considered by Sitarz.
He defined
the new metric $g_{\alpha\beta}$ with $\alpha$ and $\beta$ 
running over $0,1,2,3,4$ by $g^{\alpha\beta}={\rm diag}(+,-,-,-,-)$.
The fifth index represents the discrete space $Z_2$. Then, 
$dx^\alpha=(dx^0,dx^1,dx^2,dx^3,\chi)$ is followed.
The generalized field strength $ F(x,y)$ in Eq.(\ref{2.21})
is written by $F(x,y)=F_{\alpha\beta}(x,y)dx^\alpha\wedge dx^\beta$ where
$F_{\alpha\beta}(x,y)$ is denoted in Eq.(\ref{2.22}). Then, it is
easily derived that ${\rm Tr}\{g^{\alpha\beta}F_{\alpha\beta}(x,y)\}$
is gauge invariant. Thus, the term
\begin{equation}
|{\rm Tr}\{g^{\alpha\beta}F_{\alpha\beta}(x,y)\}|^2
=\{{\rm Tr}V(x,y)\}^\dagger\{{\rm Tr} V(x,y)\}    \label{4.1}
\end{equation}
can be added to Eq.(\ref{3.20}). This procedure gives rise to the
Lagrangian without any restriction on Higgs quartic coupling constants.
In this article, we assume that this Sitarz term vanishes at 
a point that Eqs.(\ref{3.25})-(\ref{3.27}) hold and 
perform the renormalization group analysis by adopting
the mass relation $m_{H}=\sqrt{2}m_{W}$ as an initial condition 
which holds at a renormalization point giving $g_+=g_-$ in Eq.(\ref{3.25}).
In this context, the way to obtain the BRST invariant Lagrangian
of spontaneously broken gauge theory in NCG has already presented by
Ref.\cite{LHN} and \cite{OBRST}, which includes the Fadeef-Popov ghosts
and Nakanishi-Lautrup field for the gauge fixing. Thus, we can obtain
the $\beta$ functions for various coupling constants in the standard model
just as already done.\cite{AMV}
\par
Shinohara, Nishida, Tanaka and Sogami \cite{SS} have estimated 
the Higgs boson mass 
by changing the renormalization point $\mu_0$ at which the mass relation
$m_{H}=\sqrt{2}m_{top}$ holds and obtained the result that 
$m_{H}$ approaches to $m_{top}$ from the upper side 
for large values  of $\mu_0$.
Following Shinohara and co-workers, the RG equations 
for coupling constants and the vacuum expectation value 
of the Higgs field are used to estimate the physical Higgs boson mass.
In our case, the renormalization point at which the mass relation
$m_{H}=\sqrt{2}m_{W}$ holds is determined by looking for the point
giving the relation $g^2=(5/3)g'^2$ 
which results in $\sin^2\theta_{W}=3/8$.
\par
Let us start by writing the RG equations for 
coupling constants in the standard model.
With notations 
\begin{equation}
   g_3=g_c, \hskip 1cm g_2=g, \hskip 1cm g_1=\sqrt{\frac53}g'
   \label{4.2}
\end{equation}
for SU(3)$_c$, SU(2)$_{\rm L}$ and U(1)$_{\rm Y}$ gauge coupling constants,
respectively, the RG equations are written as 
\begin{equation}
      \mu\frac{\partial\alpha_i}{\partial\mu}=\beta_i, 
      \hskip 1cm \alpha_i=\frac{g_i^2}{4\pi}\hskip 1cm (i=1,2,3),
    \label{4.3}
\end{equation}
where \cite{AMV}
\begin{eqnarray}
       && \beta_1=\frac1{2\pi}\left(\frac43n_f+\frac1{10}\right)
                        \alpha_1^2, \label{4.4}\\
       && \beta_2=-\frac1{2\pi}\left(\frac{22}3-\frac3{4}n_f
                 -\frac16 \right) \alpha_2^2, \label{4.5}\\
       && \beta_3=-\frac1{2\pi}\left(\frac{33}3-\frac43n_f\right)
                        \alpha_3^2 \label{4.6}
\end{eqnarray}
with the generation number $n_f$.
The RG equation of the quartic Higgs coupling constant $\lambda$ 
is expressed as
\begin{equation}
          \mu\frac{\partial\alpha_{H}}{\partial\mu}=\beta_{H}, 
          \hskip 1cm \alpha_{H}=\frac\lambda{4\pi},\label{4.7}
\end{equation}
where \cite{AMV} 
\begin{eqnarray}
      \beta_{H}&=&\frac6\pi \alpha_{H}^2-\frac{9}{20\pi}
                      \alpha_{H}\alpha_1
                -\frac9{4\pi}\alpha_{H}\alpha_2  
                +\frac{27}{800\pi}\alpha_1^2+\frac9{80\pi}\alpha_1\alpha_2
                +\frac9{32\pi}\alpha_2^2 \nonumber\\
                &&+\frac1{4\pi^2}{\rm Tr}
                \left\{(A_e^\dagger A_e)+3(A_u^\dagger A_u)
                +3(A_d^\dagger A_d) \right\}\alpha_{H} 
                       - \frac1{32\pi^3}
                \left\{(A_e^\dagger A_e)^2+3(A_u^\dagger A_u)^2
                +3(A_d^\dagger A_d)^2 \right\}. \label{4.8}
\end{eqnarray}
$A_e$, $A_u$ and $A_d$ in Eq.(\ref{4.8}) are the Yukawa coupling
constants written in $3\times3$ matrices
for the lepton, up quark and down quark sectors, respectively.
In view of the large top quark mass in the low energy region,
we now assume the top quark dominance in the evaluation of traces 
with respect to the Yukawa coupling constants in Eq.(\ref{4.8}).
In this approximation, $\beta_{H}$ is rewritten \cite{AMV} as
\begin{eqnarray}
      \beta_{H}&=&\frac6\pi \alpha_{H}^2-\frac{9}{20\pi}
                          \alpha_{H}\alpha_1
                -\frac9{4\pi}\alpha_{H}\alpha_2
                +\frac{27}{800\pi}\alpha_1^2
                 +\frac9{80\pi}\alpha_1\alpha_2
                +\frac9{32\pi}\alpha_2^2 
                +3\frac1{\pi}\alpha_{Y}\alpha_{H} 
              -\frac3{2\pi}\alpha_{Y}^2      \label{4.9}
\end{eqnarray}
with the definition that
\begin{equation}
    \alpha_{Y}=\frac {|(A_u)_{33}|^2}{4\pi}.      \label{4.10}
\end{equation}
$\alpha_{Y}$ is subject to the following RG equation.
\begin{equation}
     \mu\frac{\partial\alpha_{Y}}{\partial\mu}=\beta_{Y},  \label{4.11}
\end{equation}
where the $\beta-$function is given as
\begin{equation}
     \beta_{Y}=\frac{\alpha_{Y}}{4\pi}\left(9\alpha_{Y}-16\alpha_3
      -\frac92\alpha_2-\frac{17}{10}\alpha_1\right).
     \label{4.12}
\end{equation}
\par
The RG equations for $\alpha_i\;(i=1,2,3)$ in Eq.(\ref{4.3}) are
analytically solved for $n_f=3$ and the solutions are given as
\begin{eqnarray}
  && \alpha_1(t)=\frac{\alpha_1(0)}
                  {1-\alpha_1(0)\displaystyle{\frac{41}{20\pi}}t}, 
                   \label{4.13}\\
  && \alpha_2(t)=\frac{\alpha_2(0)}
                  {1+\alpha_2(0)\displaystyle{\frac{19}{12\pi}}t},
                    \label{4.14}\\
  && \alpha_3(t)=\frac{\alpha_3(0)}
                  {1+\alpha_3(0)\displaystyle{\frac{7}{2\pi}}t}, 
                   \label{4.15}
\end{eqnarray}
where $t=\log(\mu/\mu^0)$.
By use of these solutions, the RG equation for $\alpha_{Y}$ can be solved 
to be
\begin{equation}
      \alpha_{Y}(t)=F(t)\left[\frac{\alpha_{Y}(0)}
      {1-\alpha_{Y}(0)\displaystyle{\frac9{4\pi}\int_0^tF(t')dt'}}\right],
      \label{4.16}
\end{equation}
where the function $F(t)$ is defined \cite{GJ} as
\begin{equation}
    F(t)=\left(\frac{\alpha_1(t)}{\alpha_1(0)}\right)^{-17/82}
         \left(\frac{\alpha_2(t)}{\alpha_2(0)}\right)^{27/38}
         \left(\frac{\alpha_3(t)}{\alpha_3(0)}\right)^{8/7}
  \label{4.17}
\end{equation}
with the condition $F(0)=1$.
\par
Masses of all particles in the standard model are introduced by the 
vacuum expectation value $v$ of the Higgs field. 
From Eqs.(\ref{3.25})-(\ref{3.27}) and (\ref{3.30})-(\ref{3.32}), 
\begin{equation}
    m_{W}^2=\frac\pi4\alpha_2v^2,\hskip 0.2cm 
    m_{Z}^2=\frac\pi4\left(\alpha_2+\frac35\alpha_1\right)v^2\hskip 0.2cm
    m_{H}^2=8\pi\alpha_{H}v^2  \label{4.18}
\end{equation}
are given by use of the vacuum expectation value $v$.
Top quark mass is also expressed with $v$ as
\begin{equation}
    m_{top}=(A_u)_{33}\frac v{\sqrt{2}}=\sqrt{2\pi\alpha_{Y}}v.\label{4.19}
\end{equation}
In order to analyze the RG evolution of these masses, we need the RG 
evolution of $v$. It is given \cite{AMV} in the following equation.
\begin{equation}
\mu\frac{\partial v}{\partial \mu}=-\frac{v}{4\pi}
             \left(3\alpha_{Y}-\frac94\alpha_2
                          -\frac9{20}\alpha_1\right)  \label{4.20}
\end{equation}
with the same approximation as in the equations for coupling constants.
It should be noted that all variables 
in Eqs.(\ref{4.18})-(\ref{4.20}) depend on $t=\log(\mu/\mu^0)$.
\par
Eq.(\ref{4.20}) is analytically solved to result in
\begin{eqnarray}
  v(t)&=&v(0)
     \left(\frac{\alpha_{Y}(t)}{\alpha_{Y}(0)}\right)^{-1/3}
     \left(\frac{\alpha_1(t)}{\alpha_1(0)}\right)^{-7/492}
     \left(\frac{\alpha_2(t)}{\alpha_2(0)}\right)^{-9/76}
     \left(\frac{\alpha_3(t)}{\alpha_3(0)}\right)^{8/21}
\label{4.21}
\end{eqnarray}
with $t=\log(\mu/\mu^0)$.
In these RG equations, we take the renormalization point $\mu_0$  to be
the neutral gauge boson mass $m_{Z}=91.187$GeV\cite{PDG}. 
Thus, we must determine
the gauge coupling constants  
in Eqs.(\ref{4.13})-(\ref{4.15})
and $v(0)$ in Eq.(\ref{4.21}) at $\mu=m_{Z}$. 
They takes the values \cite{PDG}
\begin{equation}
     \alpha_1(0)=0.017,\hskip 0.2cm \alpha_2(0)=0.034,
     \hskip 0.2cm  \alpha_3(0)=0.12 \label{4.22}
\end{equation}
and
\begin{equation}
     v(0)=\sqrt{\frac1{\sqrt{2}G_{F}}}=246{\rm GeV}. \label{4.23}
\end{equation}
According to Eq.(\ref{4.19}), the physical top quark mass $m_{top}$ 
satisfies the equation
\begin{equation}
   m_{top}=\sqrt{2\pi \alpha_{Y}(t_{top})}v(t_{top}), \label{4.24}
\end{equation}
where $t_{top}=\log(m_{top}/m_{Z})$.
This equation along with Eqs.(\ref{4.16}), (\ref{4.17}) and 
(\ref{4.22}) fixes the numerical value of 
$\alpha_{Y}(0)$ which depends on the physical mass $m_{top}$. 
\par
Let us consider to solve the RG equation in Eq.(\ref{4.7}) by adopting
the mass relation $m_{H}=\sqrt{2}m_{W}$ as 
the initial condition. From Eqs.(\ref{3.25}), (\ref{3.30}) and (\ref{3.32}), 
this mass relation leads to the relation $g_1=g_2$ in Eq.(\ref{4.2}).
Then, according to Eqs.(\ref{4.13}) and (\ref{4.14}) 
together with Eq.(\ref{4.22}), 
we can determine the renormalization point $t_0$
 at which $m_{H}=\sqrt{2}m_{W}$ holds. 
 The numerical value of $t_0$ is
\begin{equation}
t_0=25.431. \label{4.25}
\end{equation}
From Eq.(\ref{4.18}), the initial condition of Eq.(\ref{4.7}) 
is determined as
\begin{equation}
     \alpha_{H}(t_0)=\frac1{16}\alpha_2(t_0).     \label{4.26}
\end{equation}
With these considerations, we can find the running Higgs boson mass  
from Eq.(\ref{4.18}) as
\begin{equation}
          m_{H}(t)=\sqrt{8\pi\alpha_{H}(t)}v(t).  \label{4.27}
\end{equation}
The physical Higgs boson mass $m_{H}$ is given 
by imposing the condition that
\begin{equation}
        m_{H}=\sqrt{8\pi\alpha_{H}(t_{H})}v(t_{H}) 
         \label{4.28}
\end{equation}
with $t_{H}=\log(m_{H}/m_{Z})$.
\par
Top quark mass $m_{top}$  reported by particle data
 group  \cite{PDG} is given as
\begin{equation}
     m_{top}=180 \pm 12 {\rm GeV}. \label{4.29}
\end{equation}
We investigate the Higgs boson mass by varying the top quark mass 
in the range of Eq.(\ref{4.29}).
Fig.1 shows the running Higgs boson 
mass for the top quark mass 180GeV.
The intersection of two functions $y=m_{_H}(t)$ and $y=m_{_Z}\exp{(t)}$ 
indicates the position of the physical Higgs boson mass $m_{_H}$,
which gives $m_{_H}=171.92$GeV.
Fig.2 shows the relation of the Higgs boson and top quark masses
in which the top quark mass varies from 168GeV to 192GeV according to 
the experimental data in Eq.(\ref{4.29}). Corresponding to the variation of 
top quark mass, the Higgs boson mass varies from 153.42GeV to 191.94GeV.
\par  
\begin{minipage}[t]{7cm}
    \epsfig{file=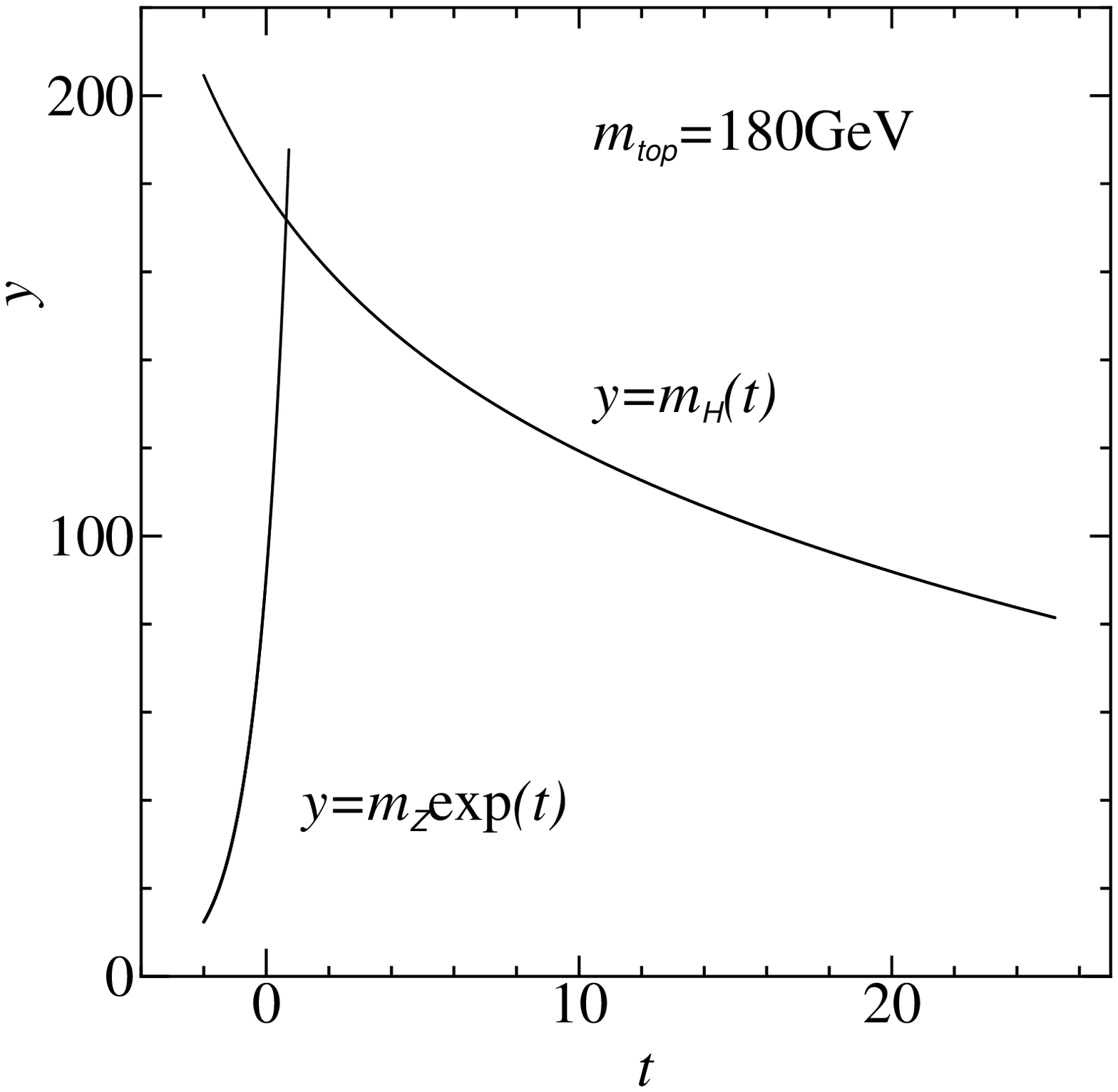,height=7cm, width=7cm}
  \vskip 0.2cm
  {Fig.1: $t=\log(\mu/m_{_Z})$ and a curve of $y=m_{_H}(t)$ shows the
  running Higgs boson mass numerically calculated according to Eq.(4.27).
  An intersection of two curves indicates the position of the physical 
  Higgs boson mass from which we can read $m_{_H}=171.92$GeV.}
\end{minipage}
\hskip 1cm
\begin{minipage}[t]{7cm}
 \epsfig{file=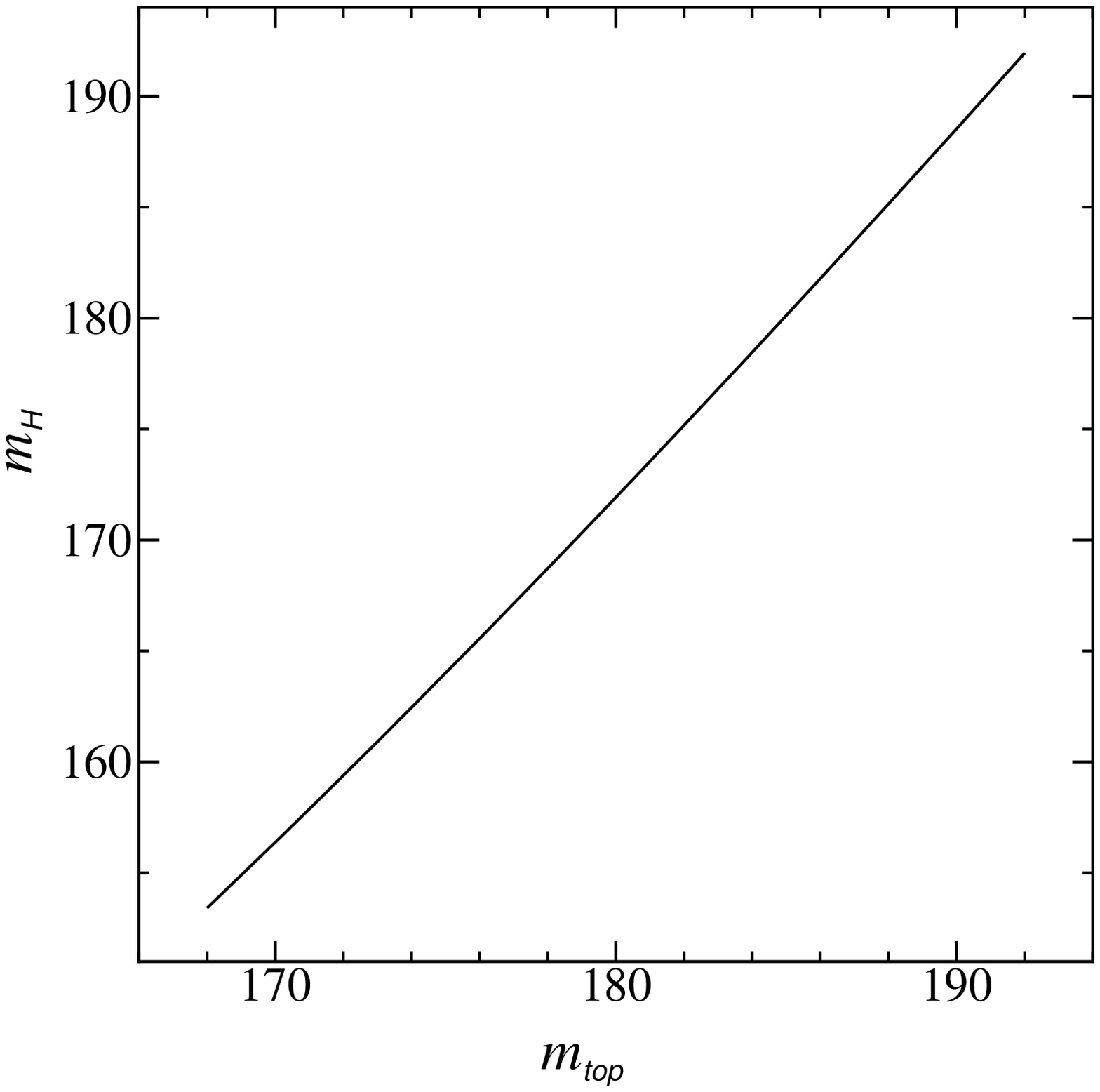,height=7cm, width=7cm}
  \vskip 0.2cm
  {Fig.2: This figure shows the relationship 
  between the Higgs boson and top quark 
  masses. Corresponding to the variation of top quark mass, 
  the Higgs boson mass varies from 153.42GeV to 191.94GeV.}
\end{minipage}
\vskip 0.5cm
\par  
 %%%%%%%%%%%%%%%%%%%%%%%%%%%%%%%%%%%%%%%%%%%%%%%%%%%%%%%%%%%%%%%%%%%%%%%%%%%%
\section{Concluding remarks}
The RG analysis has been carried out for the Higgs boson mass 
by adopting  the initial condition induced by 
$m_{H}=\sqrt{2}m_{W}$ which was presented  
in the new scheme of NCG \cite{OFS} 
for the reconstruction of the standard model. 
This mass relation together with $\sin^2\theta_{_W}=3/8$
holds under same condition 
that $g_+=g_-$ (see Eqs.(\ref{3.25a}) and (\ref{3.25})
which yields   $g^2=(5/3)g'^2$ with the SU(2)$_{_L}$ and
  U(1)$_{_Y}$ gauge coupling constants.
 However, it should be noted 
 that the color gauge coupling constant $g_c$ includes
 an additional parameter $g_s$ as in Eq.(\ref{3.27}, and so 
 $g_c$ does not necessarily coincide with the SU(2)$_{_L}$ 
 gauge coupling constant $g$. 
 Thus, contrary to SU(5) GUT without supersymmetry , 
 the grand unification of coupling constants is not realized in the present 
 scheme. In solving  the RG equation 
 we use these relations as initial conditions 
 not inconsistent with the current observation of the grand unification
 of coupling constants.\par
The numerical mass of the Higgs boson considerably 
depends on the top quark mass since the top quark Yukawa coupling
constants $\alpha_t$ gives a negative contribution to $\beta_{H}$
in the range of $-2.0\leq t \leq 25$  in Eq.(\ref{4.9}).
We have the initial condition induced by $m_{H}=\sqrt{2}m_{W}$ 
at $t=25.431$ and the calculation has performed down to $t=-2$.
The larger top quark mass is, the larger Higgs boson mass we have as in
Fig.2. If we adopt 180GeV as top quark mass, the Higgs boson mass
becomes 171.92GeV which is smaller than that predicted by 
Shinohara and co-worker. If we use the data reported by Dawson \cite{DAW},
$m_{_H}=164.01{\rm GeV}$ for $m_{top}=175{\rm GeV}$.
Generally speaking, the NCG approach provides the Yang-Mills-Higgs
Lagrangian with the restrictions on  
gauge and Higgs quartic coupling constants. Our mass relation
$m_{H}=\sqrt{2}m_{W}$ is with respect to gauge and Higgs boson masses, 
whereas  other version of NCG \cite{KDC}, \cite{AGM}
including Sogami's approach \cite{Soga} present the mass relation
between the Higgs boson and top quark.
This is because the Yukawa coupling 
constants written in matrix form in the generation space are
not contained in the generalized gauge field in our formulation.
This aspect is a characteristic point of our approach compared to others.
This is due to the introduction of one-form basis $d^\mu$ in $M_4$ and $\chi$
in $Z_2$ instead of $\gamma^\mu$ and $\gamma^5$. This leads to the meaningful
Higgs potential even in one fermion generation and 
the mass relation for the Higgs and gauge bosons.
Therefore, we predict 
relatively small Higgs boson mass as seen in this article.
\par
  Calculations have performed within one loop
approximation and top quark dominance is assumed in the
whole range of $t=\log(\mu/m_{Z})$ in the RG equations for
$\beta$ functions. Though there are fairly uncertainties 
with respect to Yukawa coupling constants for  three generations 
of fermions,
it is very interesting to analyze the RG equations
for the Higgs boson mass in two loop approximation.
This will be attempted in future work.

\vskip 0.5 cm
%%%%%%%%%%%%%%%%%%%%%%%%%%%%%%%%%%%%%%%%%%%%%%%%%%%%%%%%%%%%%%%%%%%%%%%%%%%%
\begin{center}
{\bf Acknowledgement}
\end{center}
The author would like to
express their sincere thanks to
 J.~IIzuka,
 H.~Kase, K. Morita and M.~Tanaka 
for useful suggestion and
invaluable discussions on the non-commutative geometry.
%%%%%%%%%%%%%%%%%%%%%%%%%%%%%%%%%%%%%%%%%%%%%%%%%%%%%%%%%%%%%%%%%%%%%%%%%%%%%%%
\def\jmp{J.~Math.~Phys.$\,$}
\def\pl{Phys. Lett.$\,$ }
\def\np{Nucl. Phys.$\,$}
\def\ptp{Prog. Theor. Phys.$\,$}
\def\prl{Phys. Rev. Lett.$\,$}
\def\pr{Phys. Rev. D$\,$}
\def\mp{Int. Journ. Mod. Phys.$\,$ }
%%%%%%%%%%%%%%%%%%%%%%%%%%%%%%%%%%%%%%%%%%%%%%%%%%%%%%%%%%%%%%%%%%%%%%%%%%%%%%


\begin{thebibliography}{99}
\bibitem{Con}
A.~Connes, p.9 in {\it The Interface of Mathematics and Particle
Physics}, \hfill\break
ed. D.~G.~Quillen, G.~B.~Segal, and Tsou.~S.~T.,
Clarendon Press, Oxford, 1990. See also,
Alain Connes and J. Lott, 
Nucl. Phys. {\bf B} (Proc. Suppl.) {\bf 18B}, 57 (1990).
\bibitem{MM} References in 
J.~Madore and J.~Mourad, \lq\lq Noncommutative Kaluza-Klein Theory
", hep-th/9601169.
\bibitem{Cham} 
A.~H.~Chamseddine, G.~Felder and J.~Fr\"olich,
{Phys. Lett.$\,$} {\bf B296} (1992), 109; \ Nucl.~Phys.$\,$ {\bf B395},
672(1993);\ 
A.~H.~Chamseddine and J.~Fr\"olich,
Phys.~Rev.~D$\,$ {\bf 50}, 2893 (1994). 
\bibitem{KDC}
D.~Kastler, Rev. Math. Phys. {\bf 5}, 477 (1993); 
M.~Dubois-Violette, Class. Quantum. Grav. {\bf 6}, 1709 (1989); 
R.~Coquereaux, G.~Esposito-Farese, and G.~Vaillant, Nucl.~Phys.$\,$ 
 {\bf B353}, 689 (1991); 
 M.~Dubois-Violette, R.~Kerner, and J.~Madore, J. Math.
Phys. {\bf 31}, 316 (1990);
  B.~Balakrishna, F.~G{\"u}rsey and K.C.~Wali, {Phys. Lett.$\,$}
   {\bf B254} ,430 (1991);
  Phys.~Rev.~D$\,$ {\bf 46}, 6498 (1992),     
 R.~Coquereaux, G.~Esposito-Far${\acute {\rm e}}$se and 
  G.~Vaillant, Nucl.~Phys.$\,$ {\bf B353}, 689 (1991), 
 R.~Coquereaux, G.~Esposito-Farese and F.~Scheck, 
 Int.~Journ.~Mod.~Phys., {\bf A7}, 6555 (1992); 
  R.~Coquereaux, R.~Haussling,
  N.~Papadopoulos and F.~Scheck, {\it ibit}. {\bf 7}, 2809 (1992).
\bibitem{Sita} 
A.~Sitarz, {Phys. Lett.$\,$},\ {\bf B308}, 311 (1993), 
       Jour. Geom. Phys. {\bf 15}, 123 (1995).
\bibitem{MO1}
Y.~Okumura, 
Prog.~Theor.~Phys., {\bf 91}, 959 (1994).\\
K.~Morita and Y.~Okumura, Phys.~Rev.~D,{\bf 50}, 1016 (1994).
\bibitem{Osu5}
Y.~Okumura, Phys.~Rev.~D, {\bf 50} (1994), 1026.
\bibitem{O10} 
Y.~Okumura, 
Prog.~Theor.~Phys.,{\bf 94}, 589 (1995).
\bibitem{Olr} Y.~Okumura and M.~Morita, 
Nuovo Cimento A, {\bf 109A}, 311 (1996).
\bibitem{OFS} Y.~Okumura, 
 Prog. Theor. Phys. {\bf 95}, 969 (1996).
\bibitem{AGM} E.~Alvarez, J.M.~Gracia-Bond${\acute {\rm i}}$a 
and C.P.~Mart${\acute {\rm i}}$n,
 {Phys. Lett.$\,$}, {\bf B306}, 55 (1993); {\it ibit}. {\bf B329}, 259 (1994).
\bibitem{SS} T.~Shinohara, K.~Nishida, H.~Tanaka and I.S.~Sogami,
\lq\lq Renormalization Group Effects on the Mass Relation Predicted
by the Standard Model with Generalized Covariant Derivatives, hep-ph/9606278.
\bibitem{Soga} I.S. Sogami, 
Prog.~Theor.~Phys.$\,$$\;$ {\bf 94},117 (1995); {\it ibit}. {\bf 95} (1996).
\bibitem{LHN}
C.Y.~Lee, D.S.~Hwang and Y.~Ne'eman, 
\lq\lq BRST Quantization of Gauge Theory
in Noncommutative Geometry: Matrix Derivative Approach", hep-th/9512215.
\bibitem{OBRST} Y.~Okumura, 
%\lq\lq BRST invariant Lagrangian of spontaneously
%broken gauge theory in a noncommutative differential geometry, 
%hep-th/9603045, to appear in Phys. Rev. D.
Phys. Rev. {\bf D54}, 4114 (1996).
\bibitem{AMV} H.~Arason, D.J.~Casta$\tilde{\rm n}$o, 
B.~Kesthelyi, S.~Mikaelian, 
E.J.~Piard, P. Ramond, Phys. Rev. D {\bf 46}, 3945 (1992); 
M.E.~Machacek and M.T.~Vaughn, Phys. Lett. {\bf 103B}, 427 (1981); 
Nucl. Phys. {\bf B222}, 83 (1983); {\it ibit}. {\bf B136}, 221 (1984); 
{\it ibit}. {\bf B249}, 70 (1985).
\bibitem{PDG} Particle Data Group, R.M.~Barnett {\it et al}., 
Phys. Rev. D, {\bf 54}, 1 (1994). 
%\bibitem{CDF}
%CDF Collaboration, F.~Abe et. al., Phys. Rev. Lett. {\bf 77}, 2626 (1995); 
%{\it ibit}. {\bf 75}, 3997 (1995).
\bibitem{GJ} G.~Jungman, Phys. Rev. D {\bf 46}, 4004 (1992).
\bibitem{DAW} S.~Dawson, \lq\lq DEF '96: 
The triumph of the standard model", hep-ph/9609340.
\end{thebibliography}
\end{document}